\documentstyle[preprint,aps]{revtex}

\begin{document}
\title{Nonperturbative Fermion-Boson Vartex Function 
in Gauge Theories}
\author{Han-xin He $^*$ \footnotetext{$^*$ E-mail: hxhe@iris.ciae.ac.cn}}
\address{China Institute of Atomic Energy, P.O.Box 275(18), Beijing 
102413, P.R.China \\
and Institute of Theoretical Physics, Academia Sinica, Beijing 100080, 
P.R.China }

\maketitle
\begin{abstract}
The nonperturbative fermion-boson vertex function in four-dimensional Abelian
gauge theories is self-consistently and exactly derived in terms of a complete set of normal (longitudinal) and transverse Ward-Takahashi relations for the fermion-boson and the axial-vector vertices 
in the case of massless fermion, in which the possible quantum anomalies and perturbative corrections are taken into account simultaneously.
We find that this nonperturbative fermion-boson vertex function is expressed nonperturbatively in terms of the full fermion propagator and contains the contributions of the transverse axial anomaly and perturbative corrections. The result that the transverse axial anomaly contributes to the nonperturbative fermion-boson vertex arises from the coupling between the fermion-boson and the axial-vector vertices through the transverse Ward-Takahashi relations for them and is a consequence of gauge invariance.


\end{abstract}

\newpage
\section{Introduction}
It is well known that the interactions determine the structure and properties
of any theory, and the basic interactions are described by the
interaction vertices. The fermion-boson vertex in gauge theories is such basic
interaction. In perturbation theory emploied for the processes in which the interactions are weak the interaction vertices are well defined
and can be computed order by order. On the other hand, the strong interaction phenomena such as the hadronic structure at low-energy scale and the nature of confinement in QCD can only be studied by nonperturbative approach. The Dyson-Schwinger equation(DSE) formalism[1] is such an appropriate nonperturbative approach, which provides a natural way to study the dynamical chiral symmetry breaking, confinement and other problems of hadronic physics[2]. 
In undertaking nonperturbative studies of gauge theories using DSEs, we immediately have to confront the issue of what is the nonperturbative form of the fermion-boson vertex function. This is because the DSEs are an infinite set of coupled integral equations which relate the n-point Green function to the 
(n+1)-point function; at its simplest, propagators are related to three-point
vertices. Therefore, we must find an adequite scheme in a gauge-invariant manner to truncate consistently this set of equations so that we are able to actually solve the DSEs. Finding the nonperturbative form of the fermion-boson vertex is needed to accomplish such truncation of the infinite set of equations, as well as is responsible for determining the nonperturbative behavior of the propagators.
Indeed, recent studies utilising the DSE approach show that the chiral symmetry restoration transition is strongly dependent on the nonperturbative structure of the interaction kernel in the infrared region[3]. 

Naturally, the basic approach to find the nonperturbative fermion-boson vertex is to use the Ward-Takahashi(WT) relations[4] since the normal WT relation for the
vertex relates the fermion-boson vertex (i.e. the vector vertex) to the full fermion propagetor, which is satisfied perturbatively as well as nonperturbatively. But the normal WT relation specifies only the longitudinal part of the vertex, leaving the transverse part unconstrained. How to solve exactly the transverse part of the vertex and thereby the full vertex function then becomes a crucial problem [2,5]. 
Although there have been considerable efforts to construct the transverse part
of the vertex in terms of an ansatz which satisfies some constraints[5],
however, all such attempts remain {\sl ad hoc} [2]. These constraints may serve as a guide to possible structure of nonperturbative vertex but can not derive the true nonperturbative vertex.  
Obviously the key point to obtain the complete constraints on the vertex 
must be to study the WT type constraint relations for the transverse part
of the vertex, i.e., the transverse WT relations[6,7,8], since the longitudinal part of the vertex is specified by the normal WT relation.
In Ref.[8], we studied the transverse WT relation for the fermion-boson
vertex. It showed that in order to obtain the complete solution for the fermion-boson vertex function one needs to build simultaneously the WT relations for the axial-vector and the tensor vertex functions as well. 
Up to the effects of possible quantum anomalies and perturbative corrections, this problem was solved by He in Ref.[9]. However, the symmetry in the classical theory may be destroied in the quantum theory by the quantum anomaly arising from the fermion-loop diagram. The well-known example is the anomalous nonconservation of the four-dimensional axial current, expressed by the anomalous axial-vector divergence equation, which is known as the axial anomaly or the Adler-Bell-Jackiw(ABJ) anomaly [10]. Hence the normal WT relation for the axial-vector vertex is modified by ABJ anomaly. Recently we found the anomalous 
axial-vector curl equation besides the ABJ anomaly, which is called the
transverse axial anomaly[11]. As a consequence, the transverse WT relation for the axial-vector vertex is modified by the transverse axial anomaly.

In this Latter we further derive the nonperturbative fermion-boson vertex function in four-dimensional Abelian gauge theories by self-consistently solving
a set of normal and transverse WT relations for the vector and the axial-vector vertex functions in momentum space in the case of 
massless fermion without any ansatz, in which the possible quantum anomalies and perturbative corrections are taken into account simultaneously. 
The error arose due to the inconsistent treatment in the previous study of the 
transverse part of the fermion-boson vertex[9] is also corrected by fully
self-consistent procedure.

The result shows that the full fermion-boson vertex function is expressed
nonperturbatively in terms of the full fermion propagator and contains the contributions of perturbative corrections and the transverse axial anomaly.
Such a full fermion-boson vertex function is true nonperturbatively as well as perturbatively because it is determined completely by the longitudinal and transverse WT relations which are satisfied nonperturbatively as well as perturbatively. It is particular interesting to find that the nonperturbative fermion-boson vertex function contains the contribution of the transverse axial anomaly that appears originally in the transverse WT relation for the axial-vector vertex, which has never been obtained so far by the normal perturbative calculation for the vector vertex. This result arises from
the coupling between the vector and the axial-vector vertices through the transverse WT relations for them. 
We notice that the WT relations and the appearance of quantum anomalies
are the consequences of local gauge invariance.
Thus, the nonperturbative fermion-gauge-boson vertex function
which are deduced exactly from the WT relations maintains the 
gauge invariance.

This paper is organized as follows. In Sec.II, we present a set of normal
and transverse WT relations for the vector and the axial-vector vertex 
functions, where the possible quantum anomalies and perturbative corrections are taken into account simultaneously. Based on this set of WT relations, in sec.II we derive the nonperturbative fermion-boson vertex function in the case of massless fermion by fully self-consistent procedure. Conclusions and discussions are given in Sec.IV.

\section{Normal and Transverse Ward-Takahashi Relations For the Vector and 
the Axial-Vector Vertex Functions}

The normal (longitudinal) WT identities for the vector and the 
axial-vector vertex functions in momentum space are well-known :  
\begin{equation}
q_\mu \Gamma _V^\mu (p_1,p_2)=S_F^{-1}(p_1)-S_F^{-1}(p_2),
\end{equation}

\begin{equation}
q_\mu \Gamma _A^\mu (p_1,p_2)=S_F^{-1}(p_1)\gamma _5+\gamma _5S_F^{-1}(p_2)
+2m\Gamma _5(p_1,p_2) + i \frac{g^2}{16\pi^2} F(p_1,p_2) ,
\end{equation}
where $q=p_1-p_2$, $m$ is the fermion mass, $\Gamma _V^\mu $, $\Gamma _A^\mu $, 
and $\Gamma _5$ are the vector, the axial-vector, and the pseudo-scalar vertex functions in momentum space, respectively, and $S_F(p_1)$ is the full fermion propagator. $F(p_1,p_2) $ denotes the contribution of the axial anomaly or the 
ABJ anomaly[10] in momentum space, and is defined by
\begin{eqnarray}
& &\int d^{4}x d^{4}x_{1} d^{4}x_{2} e^{i(p_{1}\cdot  x_{1} -
p_{2}\cdot  x_{2} - q\cdot  x)} \langle 0|T \psi(x_{1})
\bar{\psi}(x_{2})\varepsilon^{\mu \nu \rho \sigma} F_{\mu \nu}(x)
F_{\rho \sigma}(x) |0\rangle \nonumber \\
&=& (2 \pi)^{4} \delta^{4}(p_{1} - p_{2} - q) iS_{F}(p_{1})
F(p_{1}, p_{2})iS_{F}(p_{2}) .
\end{eqnarray}
The normal WT identities (1) and (2) specify only the 
longitudinal parts of the vector and the axial-vector vertices, respectively,
leaving the transverse parts unconstrained. In coordinate space  
they are related to the divergence of the time-ordered products of 
the three-point Green functions involving the corresponding current operators. 

In the canonical field theory the transverse WT relations for the vector and 
the axial-vector vertices in coordinate space are related to the curl 
of the time-ordered products of the three-point functions involving the
vector and the axial-vector current operators, respectively[8,9]. 
The results are

\begin{eqnarray}
& &\partial _x^\mu \left\langle 0\left| Tj^\nu (x)\psi (x_1)\bar{\psi}(x_2)
\right| 0\right\rangle -\partial _x^\nu \left\langle 0\left| 
Tj^\mu (x)\psi (x_1)\bar{\psi}(x_2)\right| 0\right\rangle \nonumber \\
&=&i\sigma ^{\mu \nu }\left\langle 0\left| T\psi (x_1)\bar{\psi}(x_2)
\right| 0\right\rangle \delta ^4(x_1-x)+i\left\langle 0\left| T\psi (x_1)
\bar{\psi}(x_2)\right| 0\right\rangle \sigma ^{\mu \nu }\delta ^4(x_2-x)\nonumber \\
& &+2m\left\langle 0\left| T\bar{\psi}(x)\sigma ^{\mu \nu }\psi (x)\psi (x_1)
\bar{\psi}(x_2)\right| 0\right\rangle \nonumber \\
& &+{\lim _{x^{\prime }\rightarrow x}}i(\partial _\lambda ^x-
\partial _\lambda ^{x^{\prime }})\varepsilon ^{\lambda \mu \nu \rho }
\left\langle 0\left| T\bar{\psi}(x^{\prime })\gamma _\rho \gamma _5
U_P (x^{\prime },x)\psi (x)\psi (x_1)\bar{\psi}(x_2)\right| 0\right\rangle ,
\end{eqnarray}
and
\begin{eqnarray}
& &\partial _x^\mu \left\langle 0\left| Tj_5^\nu (x)\psi (x_1)\bar{\psi}(x_2)
\right| 0\right\rangle -\partial _x^\nu \left\langle 0\left| Tj_5^\mu (x)
\psi (x_1)\bar{\psi}(x_2)\right| 0\right\rangle 
\nonumber \\
&=&i\sigma ^{\mu \nu }\gamma _5\left\langle 0\left| T\psi (x_1)\bar{\psi}(x_2)
\right| 0\right\rangle \delta ^4(x_1-x)-i\left\langle 0\left| T\psi (x_1)
\bar{\psi}(x_2)\right| 0\right\rangle \sigma ^{\mu \nu }\gamma _5
\delta ^4(x_2-x) \nonumber \\
& &+{\lim _{x^{\prime }\rightarrow x}}i(\partial _\lambda ^x-
\partial _\lambda ^{x^{\prime }})\varepsilon ^{\lambda \mu \nu \rho }
\left\langle 0\left| T\bar{\psi}(x^{\prime })\gamma _\rho U_P (x^{\prime },x)
\psi (x)\psi (x_1)\bar{\psi}(x_2)
\right| 0\right\rangle \nonumber \\
& &+\frac{g^2}{16\pi^2} \left\langle 0\left| T \psi (x_1)\bar{\psi}(x_2)
[ \varepsilon ^{\alpha \beta \mu \rho } F_{\alpha \beta} (x) F^{\nu}_{\rho}(x) 
- \varepsilon ^{\alpha \beta \nu \rho } F_{\alpha \beta} (x) 
F^{\mu}_{\rho}(x) ]\right| 0\right\rangle ,
\end{eqnarray}
where
$j^{\mu}(x)=\bar{\psi}(x)\gamma ^{\mu }\psi (x)$,
$j^{\mu}_{5}(x)=\bar{\psi}(x)\gamma ^{\mu} \gamma_{5} \psi (x)$,
 $\sigma ^{\mu \nu }=\frac{i}{2}[\gamma ^{\mu },\gamma ^{\nu }]$.
The Wilson line
$U_P (x^{\prime },x)=P\exp (-ig\int_x^{x^{\prime }}dy^\rho A_\rho (y))$
is introduced in order that the operators be locally gauge invariant,
where $A_\mu $ is the gauge fields.
 In the QED case, $g = e$ and $A_{\rho}$ is the photon
field. In the QCD case, $A_{\rho} = A^{a}_{\rho} T^{a}$,
$A_{\rho}^{a}$ are the non-Abelian gluon fields and $T^{a}$ are the
generators of $SU(3)_{c}$ group. 
The last term of Eq.(5) is the the contribution of the transverse axial 
anomaly expressed by the anomalous axial-vector curl equation[11]

\begin{eqnarray}
& &\partial ^\mu j_5^\nu (x) -\partial ^\nu j_5^\mu (x)
\nonumber \\
&=& {\lim _{x^{\prime }\rightarrow x}} i(\partial _\lambda ^x-
\partial _\lambda ^{x^{\prime }})\varepsilon ^{\lambda \mu \nu \rho }
\bar{\psi}(x^{\prime })\gamma _\rho U_P (x^{\prime },x)\psi (x) \nonumber \\
& & + \frac{ g^2}{16\pi^2} [ \varepsilon ^{\alpha \beta \mu \rho } 
F_{\alpha \beta} (x) F^{\nu}_{\rho}(x) - \varepsilon ^{\alpha \beta \nu \rho }
 F_{\alpha \beta} (x) F^{\mu}_{\rho}(x) ] ,
\end{eqnarray}
where the last term is the transverse axial anomaly for the curl of the 
axial-vector current, which is different from the ABJ 
anomaly expressed by the anomalous axial-vector divergence equation[10].
 
The transverse W-T relations for the vector and the axial-vector vertex
functions can be written in more clear and elegant form in 
 momentum space by computing the Fourier transformation of Eqs.(4) and
(5). In terms of the standard definition for the Fourier transformation of the three-point functions, we obtain
\begin{eqnarray}
& &iq^\mu \Gamma _V^\nu (p_1,p_2)-iq^\nu \Gamma _V^\mu (p_1,p_2)\nonumber \\
&=&S_F^{-1}(p_1)\sigma ^{\mu \nu }+\sigma ^{\mu \nu }S_F^{-1}(p_2)+
2m\Gamma _T^{\mu \nu }(p_1,p_2)\nonumber \\
& &+(p_{1\lambda }+p_{2\lambda })\varepsilon ^{\lambda \mu \nu \rho }
\Gamma _{A\rho }(p_1,p_2)
-\int \frac{d^{4}k}{(2 \pi)^{4}}2k_{\lambda}
\varepsilon ^{\lambda \mu \nu \rho }\Gamma _{A\rho }(p_1,p_2;k),
\end{eqnarray}
and
\begin{eqnarray}
& &iq^\mu \Gamma _A^\nu (p_1,p_2)-iq^\nu \Gamma _A^\mu (p_1,p_2)\nonumber \\
&=&S_F^{-1}(p_1)\sigma ^{\mu \nu }\gamma _5-
\sigma ^{\mu \nu }\gamma _5S_F^{-1}(p_2)
+(p_{1\lambda }+p_{2\lambda })\varepsilon ^{\lambda \mu \nu \rho }
\Gamma _{V\rho }(p_1,p_2)\nonumber \\
& &-\int \frac{d^{4}k}{(2 \pi)^{4}}2k_{\lambda}
\varepsilon ^{\lambda \mu \nu \rho }\Gamma _{V\rho }(p_1,p_2;k)
+\frac{g^2}{16 \pi^2} F^{\mu \nu}_{(T)}(p_1,p_2) ,
\end{eqnarray}
where $q=p_1-p_2$, $\Gamma _T^{\mu \nu }$ 
is the tensor vertex function in momentum space.
$F^{\mu \nu}_{(T)}(p_1,p_2)$ denotes the contribution of the transverse
axial anomaly in momentum space, and is defined by
\begin{eqnarray}
& &\int d^{4}x d^{4}x_{1} d^{4}x_{2} e^{i(p_{1}\cdot  x_{1} -
p_{2}\cdot  x_{2} - q\cdot  x)} \langle 0|T \psi(x_{1})
\bar{\psi}(x_{2}) [\varepsilon^{\alpha \beta \mu \rho } F^{\nu}_{\rho }(x) 
- \varepsilon^{\alpha \beta \nu \rho } F^{\mu}_{ \rho }(x) ] 
F_{\alpha \beta}(x) |0\rangle \nonumber \\
&=& (2 \pi)^{4} \delta^{4}(p_{1} - p_{2} - q) iS_{F}(p_{1})
F_{(T)}^{\mu \nu}(p_{1}, p_{2})iS_{F}(p_{2}) .
\end{eqnarray}
Note that here $ F_{(T)}^{\mu \nu}(p_1,p_2)$ is equal to 
$K^{\mu \nu}(p_1,p_2)$ defined in Ref.[11].
In Eqs.(7) and (8), $\Gamma _{V\rho}(p_1,p_2;k)$ and 
$\Gamma _{A\rho}(p_1,p_2;k)$ involve the internal momentum $k$
of the gauge boson appearing in the Wilson line and hence are the higher-order 
contributions of $\Gamma _{V\rho}(p_1,p_2)$ and 
$\Gamma _{A\rho}(p_1,p_2)$, respectively. Their contributions were negelected
in Refs.[8,9]. The integrals in Eqs.(7) and (8)
can be written order by order. For example,
to the one-loop order the integral in Eq.(7) can be 
written as 
\begin{eqnarray}
& &\int \frac{d^{4}k}{(2 \pi)^{4}}2k_{\lambda}
\varepsilon ^{\lambda \mu \nu \rho }\Gamma _{A\rho }(p_1,p_2;k)\nonumber \\
&=& g^2\int \frac{d^{4}k}{(2 \pi)^{4}}2k_{\lambda}
\varepsilon ^{\lambda \mu \nu \rho }\gamma^{\alpha}
\frac{1}{\hat{p_1}-\hat{k}-m}\gamma^{\rho}\gamma_5
\frac{1}{\hat{p_2}-\hat{k}-m}\gamma^{\beta}
\frac{-i}{k^2}[g_{\alpha\beta}+(\xi-1)\frac{k_{\alpha}k_{\beta}}{k^4}]\nonumber \\
& &+g^2\int \frac{d^{4}k}{(2 \pi)^{4}}2
\varepsilon ^{\alpha \mu \nu \rho }[\gamma^{\beta}
\frac{1}{\hat{p_1}-\hat{k}-m}\gamma^{\rho}\gamma_5 + \gamma^{\rho}\gamma_5
\frac{1}{\hat{p_2}-\hat{k}-m}\gamma^{\beta}]
\frac{-i}{k^2}[g_{\alpha\beta}+(\xi-1)\frac{k_{\alpha}k_{\beta}}{k^4}],
\end{eqnarray}
where $\hat{k}=\gamma \cdot k$, and $\xi$ is the covariant gauge parameter.
The last two terms in the right-hand side of Eq.(10) are the one-loop
self-energy contributions accompanying the vertex correction.
The integral in Eq.(8) can be written similarly. Therefore,
the terms involving the integrals over the internal
momentum $k$ correspond to the perturbative corrections to the vertices.
It can be verified that the transverse WT relations (7) and (8) are satisfied at 
one-loop order in perturbation theory, and then can be further verified that they should be satisfied at each order in perturbation theory since the perturbative
correction terms in the WT relations (7) and (8) can be written order by order.

 It is easy to show that Eqs.(7) and (8) specify the transverse parts
of the vertices [8,9]. Eqs.(7) and (8) show that the transverse parts of 
the vector and the axial-vector vertex functions are coupled each other
through the transverse WT relations for them. It implies that the 
transverse parts of the vector and the axial-vector 
vertex functions are not independent of each other 
in four-dimensional spacetime. Thus, some properties of
the axial-vector vertex function, for instance, the transverse axial anomaly
that appears originally in the transverse WT relation for the axial-vector vertex, will also contribute to the vector vertex function.

Now we have the normal WT identities (1) and (2), which impose the constraints
on longitudinal parts of the vector and the axial-vector vertex functions, 
respectively, and the transverse WT relations (7) and (8), which impose the constraints on transverse parts of these vertex functions. In general, this set of WT relations
are not enough to determine the complete solutions of the vector and the axial-vector vertex functions
because this set of WT relations involve the terms $m\Gamma _S $ and
$m\Gamma _T^{\mu\nu}$ where $m$ is fermion mass. However, in the limit with
zero fermion mass, the contributions from the pseudo-scalar and the tensor
vertex functions disappear and hence Eqs.(1), (2), (7) and (8) form a complete set
of WT relations for the vector and the axial-vector vertices. Then we can solve 
this set of WT relations to obtain the full vector and the axial-vector vertex functions in terms of the fermion propagator and perturbative corrections.

\section{Nonperturbative Fermion-Boson Vertex Functions} 

We now derive the full fermion-boson vertex function $\Gamma _V^\mu $ by consistently and exactly solving the set of WT relations for the vector 
and the axial-vector vertex functions, given by Eqs.(1), (2), (7) and (8), in the case of massless fermion.
To do this, we multiply both sides of Eqs.(7) and (8) by $iq_\nu $, and then 
move the terms proportional to $q_\nu \Gamma _V^\nu $ and 
$q_\nu \Gamma _A^\nu $ into the right-hand side of the equations. 
We thus have
\begin{eqnarray}
q^2\Gamma _V^\mu (p_1,p_2)&=&q^\mu [q_\nu \Gamma _V^\nu (p_1,p_2)]+
iS_F^{-1}(p_1)q_\nu \sigma ^{\mu \nu }+iq_\nu \sigma ^{\mu \nu }S_F^{-1}(p_2)
\nonumber \\
 & &+i(p_{1\lambda }p_{2\lambda })q_\nu \varepsilon ^{\lambda \mu \nu \rho }
\Gamma _{A\rho }(p_1,p_2) 
- iq_{\nu}\int \frac{d^{4}k}{(2 \pi)^{4}}2k_{\lambda}
\varepsilon ^{\lambda \mu \nu \rho }\Gamma _{A\rho }(p_1,p_2;k),
\end{eqnarray}
\begin{eqnarray}
q^2\Gamma _A^\mu (p_1,p_2)&=&q^\mu [q_\nu \Gamma _A^\nu (p_1,p_2)]+
iS_F^{-1}(p_1)q_\nu \sigma ^{\mu \nu }\gamma _5-iq_\nu \sigma ^{\mu \nu }
\gamma _5S_F^{-1}(p_2)\nonumber \\
& &+i(p_{1\lambda }+p_{2\lambda })q_\nu \varepsilon ^{\lambda \mu \nu \rho }
\Gamma _{V\rho }(p_1,p_2) 
- iq_{\nu}\int \frac{d^{4}k}{(2 \pi)^{4}}2k_{\lambda}
\varepsilon ^{\lambda \mu \nu \rho }\Gamma _{V\rho }(p_1,p_2;k)\nonumber \\
& &+i\frac {g^2}{ 16 \pi^2} q_{\nu}F^{\mu \nu}_{(T)}(p_1,p_2).
\end{eqnarray}
By substituting Eq.(12) into Eq.(11) and using Eqs.(1) and (2), after some
lengthy computation
we obtain the following full fermion-boson vertex function in the case of massless fermion
\begin{eqnarray}
\Gamma _{V}^\mu (p_1,p_2)&=&[q^2+(p_1+p_2)^2 -
((p_1+p_2)\cdot q)^2q^{-2}]^{-1}\nonumber \\
& &\times \{[S_F^{-1}(p_1)-S_F^{-1}(p_2)][q^\mu + q^\mu (p_1+p_2)^2q^{-2}
- q^\mu ((p_1+p_2)\cdot q)^2q^{-4}]\nonumber \\
& &+iS_F^{-1}(p_1)\sigma ^{\mu \nu }q_\nu +i\sigma ^{\mu \nu }q_\nu S_F^{-1}(p_2)\nonumber \\
& &+i[S_F^{-1}(p_1)\sigma ^{\mu \lambda }-\sigma ^{\mu \lambda }S_F^{-1}(p_2)]
(p_{1\lambda }+p_{2\lambda })\nonumber \\
& &+i[S_F^{-1}(p_1)\sigma ^{\lambda \nu }-\sigma ^{\lambda \nu }S_F^{-1}(p_2)]
q_\nu (p_{1\lambda }+p_{2\lambda })q^\mu q^{-2}\nonumber \\
& &-i[S_F^{-1}(p_1)\sigma ^{\mu \nu }-\sigma ^{\mu \nu }S_F^{-1}(p_2)]
q_\nu (p_1+p_2)\cdot qq^{-2}\nonumber \\
& &+i[S_F^{-1}(p_1)\sigma ^{\lambda \nu }+\sigma ^{\lambda \nu }S_F^{-1}(p_2)]
q_\nu (p_{1\lambda }+p_{2\lambda })
[p_1^\mu +p_2^\mu -q^\mu (p_1+p_2)\cdot qq^{-2}]q^{-2}
\nonumber \\
& &+C^{\mu}_{V+A} + (p_{1\nu}+p_{2\nu})C^{\nu}_{V+A}
[p_1^\mu +p_2^\mu -q^\mu (p_1+p_2)\cdot qq^{-2}]q^{-2}\nonumber \\
& &-\frac{g^2}{16\pi^2} (p_{1 \lambda} + p_{2 \lambda}) q^{- 2} 
q_{\nu} q^{\sigma} \varepsilon^{\lambda \mu \nu \rho} 
F_{\rho \sigma}^{(T)}(p_1, p_2) \} 
\end{eqnarray}
with
\begin{eqnarray}
C^{\mu}_{V+A}&=& (p_{1\lambda }+p_{2\lambda })q_\nu q_\alpha q^{-2}
\varepsilon^{\rho \lambda \mu \nu} \varepsilon^{\rho \beta \alpha \delta} 
\int \frac{d^{4}k}{(2 \pi)^{4}}2k_{\beta}\Gamma _{V\delta}(p_1,p_2;k)
\nonumber \\
& &- iq_{\nu}\int \frac{d^{4}k}{(2 \pi)^{4}}2k_{\lambda}
\varepsilon ^{\lambda \mu \nu \rho }\Gamma _{A\rho }(p_1,p_2;k),
\end{eqnarray}
where the terms involving $C^{\mu}_{V+A}$ arise from the perturbative corrections. The explicit form of $C^{\mu}_{V+A}$ can be written after performing the loop-integrals and the expression is rather lengthy, which will be discussed elsewhere. The last term of Eq.(13) is the contribution of the transverse axial anomaly. The coupling between the vector and the axial-vector
vertices through the transverse WT relations for them, Eqs.(7) and (8),
leads to that the transverse axial anomaly which appears originally in the 
transverse WT relations for the axial-vector vertex contributes to the 
fermion-boson (vector) vertex function.
 
We can also write the full fermion-boson vertex function as a sum of its
longitudinal and transverse parts:
\begin{equation}
\Gamma _V^\mu (p_1,p_2)=\Gamma _{V(L)}^\mu (p_1,p_2)+\Gamma _{V(T)}^\mu (p_1,p_2)
\end{equation}
with
\begin{equation}
\Gamma _{V(L)}^\mu (p_1,p_2)=q^{-2}q^\mu [S_F^{-1}(p_1)-S_F^{-1}(p_2)],
\end{equation}
\begin{eqnarray}
\Gamma _{V(T)}^\mu (p_1,p_2)&=&[q^2+(p_1+p_2)^2-
((p_1+p_2)\cdot q)^2q^{-2}]^{-1}\nonumber \\
& &\times \{iS_F^{-1}(p_1)\sigma ^{\mu \nu }q_\nu 
+i\sigma ^{\mu \nu }q_\nu S_F^{-1}(p_2)\nonumber \\
& &+i[S_F^{-1}(p_1)\sigma ^{\mu \lambda }-\sigma ^{\mu \lambda }S_F^{-1}(p_2)](p_{1\lambda }+p_{2\lambda })\nonumber \\
& &+i[S_F^{-1}(p_1)\sigma ^{\lambda \nu }-\sigma ^{\lambda \nu }S_F^{-1}(p_2)]q_\nu (p_{1\lambda }+p_{2\lambda })q^\mu q^{-2}\nonumber \\
& &-i[S_F^{-1}(p_1)\sigma ^{\mu \nu }-\sigma ^{\mu \nu }S_F^{-1}(p_2)]q_\nu (p_1+p_2)\cdot qq^{-2}\nonumber \\
& & +i[S_F^{-1}(p_1)\sigma ^{\lambda \nu }+\sigma ^{\lambda \nu }S_F^{-1}(p_2)]
q_\nu (p_{1\lambda }+p_{2\lambda })
[p_1^\mu +p_2^\mu -q^\mu (p_1+p_2)\cdot qq^{-2}]q^{-2}
\nonumber \\
& &+C^{\mu}_{V+A} + (p_{1\nu}+p_{2\nu})C^{\nu}_{V+A}
[p_1^\mu +p_2^\mu -q^\mu (p_1+p_2)\cdot qq^{-2}]q^{-2}\nonumber \\
& &- \frac{g^2}{16\pi^2} (p_{1 \lambda} + p_{2 \lambda}) q^{- 2} 
q_{\nu} q^{\sigma} \varepsilon^{\lambda \mu \nu \rho} 
F_{\rho \sigma}^{(T)}(p_1, p_2) \} .
\end{eqnarray}
We note that this result improves the previous expression given by Eq.(26)
of Ref.[9] in following aspects: The contributions of the transverse
axial anomaly and the perturbative corrections denoted by $C^{\mu}_{V+A}$  
are included. Besides, in the
process of deriving Eq.(26) of Ref.[9] we used inconsistently the formula
\begin{equation}
(p_{1\mu}+p_{2\mu})\Gamma_V^{\mu}(p_1,p_2)
= 2m\Gamma_S(p_1,p_2) + S^{-1}_F(p_1) + S^{-1}_F(p_2),
\end{equation}
which led to the error in the expression of the transverse part of the fermion-boson vertex in the previous study[9].
This defect is overcomed in present work by the self-consistent iteration procedure and now the result is corrected. In this paper, Eqs.(13) and (17) are derived exactly and self-consistently by means of WT relations (1), (2), (7) and (8) without any approximation. 

The full fermion-boson vertex function now is expressed nonperturbatively in terms of the full fermion propagator and contains the contributions of the transverse axial anomaly and the perturbative corrections given by terms involving $C^{\mu}_{V+A}$.
This is the nonperturbative form of the fermion-boson vertex function because
this vertex function is completely determined in terms of a set of WT relations which are true nonperturbatively.

Above results are given in the Abelian case. In QCD case we usually consider the three-point functious involving the current operators,
$j^{\mu}(x)=\bar{\psi}(x)\gamma ^{\mu }\frac{\lambda^{a}}{2}\psi (x)$ and
$j^{\mu}_5(x)=\bar{\psi}(x)\gamma ^{\mu} \gamma_5 \frac{\lambda^{a}}{2}\psi (x)$,
respectively, where $\lambda^{a} $ are the generators of flavour SU(n) group. 
In such case, for the axial currents of non-Abelian QCD, the anomaly equation should be the Abelian result, supplemented by the appropriate group theory factors. Besides, the relative results of present work will be modified by putting the generator factors into suitable position in each term of corresponding relations. For the case of effective QCD with Faddeev-Popov ghost fields, the transverse WT type relations need to be studied further.

\section{Conclusions and Discussions}

 In this paper we have derived the nonperturbative form of the full
fermion-boson vertex function in four-dimentional Abelian gauge theories in terms of a complete set of normal (longitudinal) and transverse WT relations for the vector and the axial-vector vertex functions in the case of massless fermion, in which the possible quantum anomalies and perturbative corrections have been taken into account simultaneously. The longitudinal part of the vertex is specified nonperturbatively by the normal WT relation for the vertex in terms of the fermion propagator, while the transverse part of the vertex is completely determined by a set of transverse WT relations for the vector and the axial-vector vertex functions (the longitudinal WT relations do not contribute to the transverse part of the vertex). The error arose from the inconsistent treatment in the previous study[9] of the fermion-boson vertex has also been corrected by fully self-consistent procedure.

The result shows that the full fermion-boson vertex function is expressed 
nonperturbatively in terms of the full fermion propagator and contains the contributions of the transverse axial anomaly and perturbative corrections.
Such a full fermion-boson vertex function is true nonperturbatively as well as
perturbatively because it is determined completely by the WT relations
which are satisfied both nonperturbatively and perturbatively. It is easy to check that this vertex function has correct perturbative limit. Indeed, if the fermion propagator is taken as bare one and the effect of quantum anomaly is neglected, then the vertex given by Eq.(13) is reduced to bare one, 
$\Gamma _V^{\mu}(p_1,p_2) \rightarrow \gamma^{\mu}$.
The one-loop vertex function can be obtained by substituting the one-loop
propagator and including the contributions of integrals over $k$ shown
in Eqs.(7) and (8), which will be discussed elsewhere.

The nonperturbative fermion-boson vertex function given by Eq.(13) or Eqs.(15)-(17) is written in the general form. 
In the practical application to the DSE for the fermion propagator,
the full fermion propagator $S_F(p)$ is usully expressed nonperturbatively 
in terms of two Lorentz scalar functions, $F(p^2)$ the wavefunction renomalization and $M(p^2)$ the mass function, so that 
$S_F(p)=F(p^2)/[\gamma \cdot p-M(p^2)]$ ( $S_F(p)$ can also be written
in the way that $S_F^{-1}(p)=A(p^2)\gamma \cdot p-B(p^2)$ ).
Substituting the expression of $S_F(p)$ into Eq.(13) or Eqs.(16)-(17),
then the full fermion-boson vertex function can be expressed nonperturbatively in terms of $F(p^2)$ and $M(p^2)$ (or $A(p^2)$ and $B(p^2)$).
Ball and Chiu have shown how to construct the longitudinal part of the
vertex in a way free of kinematic singularities[5]. According to the way
discribed by Ball and Chiu, in priciple, there is no difficulty to rewrite
the transverse part of the vertex in a manner free of kinematic singularities.

We note that, as is well-known, the WT relations are essential in the 
renormalization programme of any theory. In fact, one of the goals of the 
renormalization of the theory is precisely to transform the formal WT relations
into relations among renormalized finite quantities. It has been verified[12]
that the consequence of a symmetry in terms of WT relations is consistent
with renormalization, and the renormalized Green functions satisfy the WT relations as a trivial consequence of multiplicative renormalizability.
Therefore, the fermion-boson vertex determined completely in terms of
the WT relations should be consistent with the multiplicative renormalizability. 
   
It is particular interesting to notice that the nonperturpative fermion-boson 
vertex function involves the contribution of the transverse axial anomaly that
appears originally in the transverse WT relation for the axial-vector vertex.
As we known, the normal perturbative calculation of the fermion-boson
vertex has never led to such result.
This result arises here due to the coupling between the transverse components of the vector and the axial-vector vertices through the transverse WT relations for them. 
Note that the appearance of the transverse axial anomaly and the ABJ 
anomaly is a consequence of local gauge invariance. On the other hand, it is well known that the WT relations are the fundamental consequence of local gauge invariance. Hence the appearance of the transverse axial anomaly in the nonperturbative fermion-boson vertex function is a natural consequence of gauge invariance.  Thus, applying this nonperturbative fermion-boson vertex function, which are determined completely in terms of the WT relations, to the Dyson-Schwinger equations in Abelian theories will provide an adequate scheme for truncating the infinite set of coupled DSEs in a gauge-invariant manner.
Up to the effects of quantum anomaly and the perturbative corrections given by the loop-integrals in Eqs.(7) and (8), the DSEs will form a self-consistent and
closed system for the fermion and gauge-boson propagators but the local gauge
invariance may be lost. The price of ensuring gauge invariance for truncating
the infinite set of DSEs is the anomalous breaking of the closed DSE system for the full propagators. This situation seems to be very similar to that for the axial current, in which the price of requiring gauge invariance is the anomalous nonconservation of the axial current.
This result may have an interesting implication for the DSE approach. It implies that the quantum anomaly may play a significant role in the nonperturbative studies of gauge theories using Dyson-Schwinger equation formalism, which involves a deeper aspect of the gauge theories and needs to be studied further.

\section*{Acknowledgments}

The author is very grateful to F.C.Khanna and Y.Takahashi for useful 
discussions.  
This work is supported by the National Natural Science Foundation 
of China under Grant Nos.19835010 and 10075081.

\end{document}